\documentclass[12pt,preprint]{aastex}

\lefthead{Thompson}
\righthead{Specific Star Formation Distribution Function}

\begin{document}

\title{Matching the Observed Star Formation Intensity Distribution with
Empirical Laws}

\author{Rodger I. Thompson}
\affil{Steward Observatory, University of Arizona, Tucson, AZ 85721}

\begin{abstract}

This letter matches the shape of the star formation intensity distribution
function to empirical laws such as the Schmidt law.  The shape of the
distribution at a redshift of one is reproduced from the empirical
Schmidt law with a critical density, a Schechter distribution of galaxy
masses and the assumption that star formation occurs mainly in
exponential disks.  The shape of the distribution depends primarily on
two values, the characteristic mass m$^*$ in the Schechter mass
distribution and the characteristic radius r$_e$ in the exponential
disk.  As these characteristic values evolve they will affect the shape
of the distribution function. The expected direction of evolution of
the parameters partially cancels each other leaving the distribution
shape relatively invariant.

\end{abstract}


\section{Introduction}

This letter has two goals.  The first is to determine if the observed
star formation intensity distribution is a natural consequence of
empirical laws such as the Schmidt law and the Schechter mass
function.  The second is to determine which parameters in the empirical
law most influence the shape of the distribution. Galaxy modelers can
then predict the evolution of the distribution with redshift and
compare it against future observations.  The star formation intensity
distribution developed by \citet{lanz99} has received significant
attention both as a constraint on models of galaxy formation
\citep{brk02} and as a method of correcting for star formation missed
by the effects of surface brightness diming at high redshifts
\citep{thm01,lanz02}.  At present, the distribution is only empirically
derived from the observations by fitting the distribution at low
redshift.  Correction of higher redshift observations for surface
brightness dimming is achieved by matching the bright end of the fitted
distribution to the bright end of the observations at higher redshift.
In particular \citet{thm01} used the empirical fit to the distribution
function at a redshift of one to correct observations at higher
redshifts.  In that case it was simply postulated that the shape of the
distribution function was invariant with redshift.  The purpose of this
work is to see if the shape of the distribution function is governed by
basic physical parameters and empirical laws such as the Schmidt law
\citep{ken98}.  Once the primary parameters influencing the
distribution shape are found estimates can be made on whether the
distribution is invariant or evolves with redshift.  This is
particularly important when the distribution is used to correct for the
effects of surface brightness dimming.

\section{Observations}

A combination of observations in the Northern Hubble Deep Field with
WFPC2 \citep{wil96} and with NICMOS \citep{dic00} define the empirical
form of the distribution.  These observations were analyzed in
essentially the same manner as described in \citet{thm01}.  The present
paper utilizes just the star formation intensity distribution at a
redshift of one.  The full analysis of the star formation history of
the entire Northern HDF will be published in a subsequent paper.  The
width of the z = 1 redshift bin is from z of 0.5 to 1.5, comprising 437
galaxies and $ 2.34 \times 10^6 $ pixels.  When known, spectroscopic
redshifts from \citet{chn01} replace the photometric redshifts in
determining the SFR intensity distribution.  This affects only a very
small percentage of galaxies. Fig.~\ref{fig:hx} shows the empirical
distribution function defined from the NHDF observations at z = 1.

Primarily due to variations in sensitivity over the field of the NICMOS
Camera 3 \citet{lanz02} developed a selection function measuring
the probability of a pixel with redshift z and star formation
intensity x being detected by the source extraction technique.  For
consistency only, the technique was also used in this analysis but its
effect on the data at redshift 1 is very negligible. For each pixel
detected by the source extraction program the fraction f of the
detector area that it could be detected in was calculated. If the
fraction was less than 1 the  proper area that the pixel contributed
to h(x) was increased by $1/f$. Fractions less than 0.1 were set to 0.1
even though very few detected pixels have fractions below 0.1.
Inclusion of the selection function correction only  eliminates a slope
inflection at a log(x) value of -1.5. 

\section{Computed Distribution Function}

The star formation intensity x is defined as the star formation rate in
solar masses per year per \emph{proper} square kiloparsec. The
intensity is calculated for each pixel that is part of a galaxy.
Within a given redshift interval the distribution function, h(x), is
defined as the sum of all the proper areas in a interval of star formation
intensity, divided by the interval and by the comoving volume in cubic
megaparsecs defined by the field and redshift interval \citep{lanz99}.
Defined in this manner the values of h(x) determine the star formation
rate per cubic comoving megaparsec through eqn.~\ref{eq:sfr}.

\begin{equation} sfr = \int_{0}^{\infty} xh(x)dx \label{eq:sfr}
\end{equation}

The star formation rate is then the first moment of the distribution
function.  It is important to note that the value of h(x) for any
interval of x is comprised of pixels from a very large number of
galaxies.  This attribute makes it an excellent vehicle for assessing
general laws of star formation. The smoothness of the h(x)distribution
over almost 8 decades of x is quite remarkable.  \citet{brk02} points
out that although h(x) is dependent on cosmology through the comoving
volume, x is independent of cosmology when it is calculated from the UV
surface brightness. He also points out that calculations of galaxy
formation and evolution should be constrained to match the observed
distribution.  Barkana then utilizes a hierarchical galaxy formation
and evolution code to predict the distribution and matches it with the
distribution function given in \citet{lanz02}.

\begin{figure}

\plotone{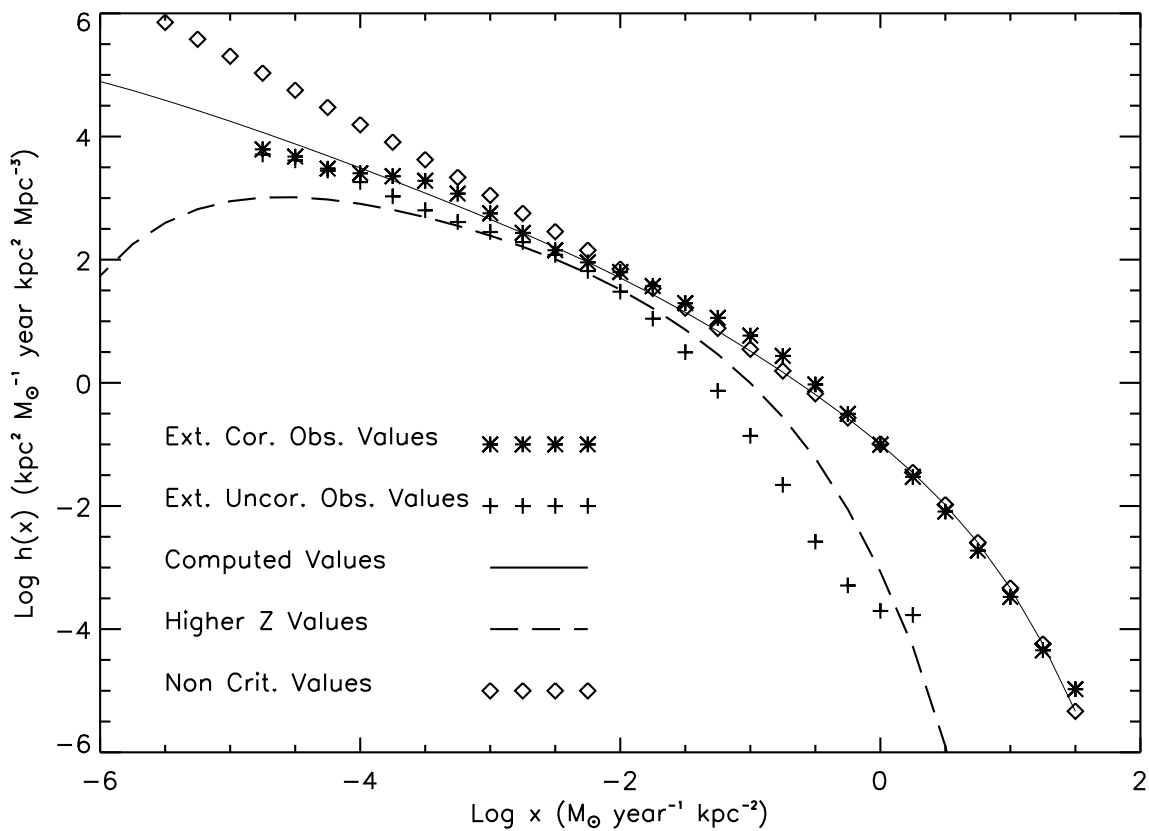}

\caption{The values of h(x) at a redshift of one from the Northern HDF
in logarithmic intervals of 0.25 in x.  The extinction corrected
observed values are denoted by * and observed values not corrected for
extinction by plus signs.  The solid line is the computed fit to the
function as described in the text. The dashed line is the expected
evolution of the distribution to an epoch when $m^*$ is 1/10 the value
at z = 1.  The diamonds show the fit if a critical density for star
formation is not invoked.}

\label{fig:hx}

\end{figure}

\placefigure{fig:hx}

It is important to note that the distribution function, h(x), in
fig.~\ref{fig:hx} utilizes \emph{extinction corrected}
star formation intensity values and is therefore different than the
distribution functions in \citet{lanz02} which have not been corrected
for extinction.  The extinction corrected distribution reflects the
basic physical parameters of the galaxies rather than the variance of
extinction values.  \citet{thm01} give a detailed account of the star
formation rate calculation and extinction correction. Briefly the 6
galaxy fluxes in a $0.6 \arcsec$ diameter aperture are  matched via
chi-squared analysis against SED template fluxes that have  been
numerically redshifted and extincted by the obscuration law of
\citet{cal94} to give a single extinction for the galaxy. The total
star formation rate for the galaxy is calculated from the 1500 \AA \space 
flux of the selected template with zero extinction, giving the
extinction corrected star formation rate.  The star formation rate
assigned to a pixel in a galaxy is the total star formation rate for
the galaxy times the fraction of extinction corrected 1.6 $\micron$
total luminosity represented by the pixel. 1.6 $\micron$ was picked
because it is the wavelength least affected by dust.

The extinction correction is the major correction to the observed
data.  Extinction correction moves pixels horizontally to the right in
figure~\ref{fig:hx} as can be seen by the difference between the
corrected and uncorrected points.  The corrected h(x) must be a
\emph{lower limit} on the true h(x) at low values of x for two
reasons.  First, pixels in a galaxy that are below our detection limit
due to extinction will not be included and second, entire galaxies that
fall below our detection limit due to high extinction values will not
be included.  The break in the corrected data at log(x) = -3.5
is most probably due to the first effect, although failure of the
selection function at this level may also contribute.  The faintest
detected pixels lie at log(x) = -5.0, 30 times fainter than the break
value.  The difference between the faintest and the break value
corresponds to an E(B-V) value of 0.4 which is toward the higher end
of the extinctions found for the sample.  Values of h(x) past log(x) =
-3.5 should not be considered as accurate, however ,the character of
both the power law slope and the exponential fall off are well
established by the distribution at log(x) values greater than -3.5.

The observed distribution has two main components, a power law
component at low x values and an exponential fall off at high x
values.  This Schechter like distribution \citep{sch74} is very common
in astronomical phenomena.  If either the power law were extended to
high values of x, or the exponential to low values of x, the total star
formation rate would diverge.  Given this familiar form of the
distribution it is interesting to see if it arises naturally from known
general empirical laws regarding mass distribution in galaxies and
relations between gas density and star formation.  Formulation in terms
of these general relations can also give insight into whether and how
the distribution function might evolve with redshift.

The derivation of the distribution shape utilizes two general empirical
laws and an assumption.  The two empirical laws are the Schmidt law
with a critical density relating star formation intensity to gas
surface density and a Schechter law for the distribution of galaxy
masses.  The assumption is that star formation occurs predominantly in
exponential disks. The following uses a dimensionless mass variable $y
= m/m^*$ where $m^*$ is the mass parameter in the Schechter mass
distribution.  $\phi^*$ is given in $m^*$ per comoving Mpc$^3$ in
eqn.~\ref{eqn:shc}.  In the following mass always refers to the gas
mass in the galaxy since we are computing star formation.

\begin{equation} \phi(y) = \phi^*y^{\alpha}\exp{(-y)} \label{eqn:shc}
\end{equation}

The formulation of the Schmidt Law comes from \citet{ken98} eqn. 7.

\begin{equation} \Sigma_{sfr} = a_o(\frac{m^*}{10^6}\Sigma_{gas})^{1.4} = 
a'_o\Sigma_{gas}^{1.4} \label{eqn:slaw} \end{equation}

\noindent
where a$_o = 2.5 \times 10^{-4}$ in the appropriate units and m$^*$ is
in solar masses. The extra term $(\frac{m^*}{10^6})^{1.4}$ comes from
working in mass units of y and gas densities in terms of square kpc.
Studies \citep{ken89,mar01} indicate that the Schmidt law does not
extend to very low densities.  There appears to be a critical density
below which star formation is severely curtailed.  The critical density
is a function of the dynamics of the system and therefore not a
constant density for all galaxies.  To account for this phenomena we
multiply the star formation efficiency $a'$ by a function of the form

\begin{equation} a' = a'_o(1+\frac{.01}{x})^{-1.5} \label{eqn:crit}
\end{equation}

\noindent 
reducing the star formation rate at low star formation intensities
which is equivalent to low surface densities. From eqn.~\ref{eqn:slaw} a
star formation intensity of 0.01 M$_{\odot}$ per year per kpc$^2$
corresponds to a surface density of 14 M$_{\odot}$ per pc$^2$ which
marks the surface density where the star formation rate begins to
deviate from the Schmidt law in this formulation.  The critical density
is discussed further in sec.~\ref{sec:crit}

For a galaxy of mass y in units of $m^*$ the assumption of an
exponential disk with a characteristic radius $r_e$ gives a gas surface
density as shown in eqn.~\ref{eqn:den}.

\begin{equation} \Sigma_{gas}(r) = \Sigma_o \exp \left(\frac{-r}{r_{e}}\right) = 
\frac{y}{2 \pi r_{e}^{2}} \exp \left(\frac{-r}{r_{e}}\right) \label{eqn:den} 
\end{equation}

\noindent
where

\begin{equation} 
r_e = r_o y^{\frac{1}{n}} \label{eqn:re}
\end{equation}

\noindent
Setting the area integral of the density equal to y determines the
value of $\Sigma_o$.  In~\ref{eqn:re} both $r_o$ and n are adjustable
variables. The final values of $r_o = 1.4$ kpc and n = 5 appear to be
within a reasonable range.

For a galaxy with a mass of y in units of $m^*$ each radius r
corresponds to a particular star formation intensity x.

\begin{equation} \left(\frac{y^{1-\frac{2}{n}}}{2 \pi r_{o}^{2}} \exp 
\left(\frac{-r}{r_{o}y^{\frac{1}{n}}}\right)\right)^{1.4} 
= \frac{x}{a'} \label{eqn:rx}
\end{equation}

\noindent This leads to an equation for the radius r(x,y) at which the sfr
intensity is x in a galaxy of mass y.

\begin{equation} r(x,y) = r_oy^{\frac{1}{n}}((1- \frac{2}{n})\ln (y) - K(x)) 
\label{eqn:rxy}
\end{equation}

\noindent 
where K(x) is given by

\begin{equation} K(x) = \ln (2\pi{r_o}^2) + \frac{1}{1.4}\ln (\frac{x}{a'})
\label{eqn:kx}
\end{equation}

\noindent 
The simple requirement that the radius r(x,y) be greater than or equal
to 0 puts a lower limit on the mass y of a galaxy that can contribute
to a star formation intensity x.  That minimum mass $y_m$ is given by

\begin{equation} y^{1-\frac{2}{n}}_m = 2 \pi r_o^2(\frac{x}{a'})^{1.4} 
\label{eqn:xm}
\end{equation}

A given star formation intensity interval $\Delta x$ corresponds to a
radius interval $\Delta r$ in a galaxy equal to $\Delta x
\frac{dr}{dx}$. The 0.25 logarithmic intervals of x used in
fig~\ref{fig:hx} correspond to an interval of 0.584x which is a $\Delta
r$ of $\frac{0.584}{1.4} r_e$.  The area corresponding to intensity
interval is simply $2 \pi r(x,y) \Delta r$.  The total area for the
intensity interval is obtained by integrating over all masses y
weighted by the probability of galaxies with that mass given by
eqn.~\ref{eqn:shc}.

\begin{equation} 
area(x) = 2\pi\frac{0.584}{1.4} r_{o}^{2}\phi^{*} 
\int_{y_m}^{\infty}((1-\frac{2}{n})\ln y - K(x))y^{\alpha + \frac{2}{n}} 
\exp(-y) dy \label{eqn:area}
\end{equation}

Integration of the second term of eqn.~\ref{eqn:area} yields K(x) times
the incomplete gamma function $\Gamma(\alpha +1+\frac{2}{n},y_m))$.
The integration of the first term was performed numerically with
Mathematica.  The area is then

\begin{equation} area(x) = 2\pi\frac{0.584}{1.4} r_{o}^{2}\phi^{*} 
(\int_{y_m}^{\infty} ((1-\frac{2}{n})y^{\alpha + \frac{2}{n}} \ln y  
\exp (-y)) dy - K(x) \Gamma(\alpha + 1 + \frac{2}{n},y_m)) \label{eqn:ar}
\end{equation}

Equation~\ref{eqn:ar} is appropriate for a galaxy viewed face on but
the observations include galaxies at all inclinations.  Inclination of
a galaxy by an angle $\theta$ to the line of sight increases the
observed x value by $(\cos(\theta))^{-1}$ over the face on x value.
This has three effects on equation~\ref{eqn:ar}.  The value of x in
K(x) should be $x \cos(\theta)$ since a value $x \cos(\theta)$ in the
face on frame appears to be x in the observed frame.  Next the radius
$\Delta r$ of the ring spanning $\Delta x$ is now $\Delta r
\cos(\theta)$ since $\Delta x$ is now $\Delta x \cos(\theta)$ in the
face on coordinates.  Finally the projected area is also reduced by
$\cos(\theta)$.  This transforms equation~\ref{eqn:ar} to

\begin{equation} area(x) = 2\pi\frac{0.584 (\cos(\theta))^2}{1.4} 
r_{o}^{2}\phi^{*} 
(\int_{y_m}^{\infty} ((1-\frac{2}{n})y^{\alpha + \frac{2}{n}} \ln y  
\exp (-y)) dy - K(\cos(\theta) x) \Gamma(\alpha + 1 + \frac{2}{n},y_m)) 
\label{eqn:arth}
\end{equation} 

Under the assumption that all inclinations are equally probable the
range between 0 and 90$^{\circ}$ is divided into 100 equally spaced
angles and equation~\ref{eqn:arth} is summed over all angles.  The last
20 angles in the angle distribution are set equal to the 80th value to
recognize the galaxies have significant thickness which also avoids the
singularity at 90$^{\circ}$ inclination.

The areas found for each intensity value are divided by the intensity
interval to determine the calculated h(x) shown as the solid line in
fig.~\ref{fig:hx}.  Free parameter values of $\phi^* = 0.007 m^*$ per
comoving cubic Megaparsec, m* = $4 \times 10^{10}$ M$_{\odot}$, $r_o$ =
2.0 kpc, $\alpha$ = -1.2 and n = 5 produce a good fit to the observed
h(x). It could be asked whether with 5 free parameters can you always
get a good fit?  On the other hand the empirical laws governing the
distribution contain these parameters and their values must be set.
That the values are consistent with values that one would set a priori
in an attempt to model the distributions gives a good indication that
the distribution shape is determined by the empirical laws and
assumptions employed.  Experiments setting one of the parameters to a
mildly nonphysical value indicated that no rearrangement of the
remaining parameters could produce an acceptable fit.

\section{Critical Density} \label{sec:crit}

The diamonds in fig.~\ref{fig:hx} indicate the computed distribution if
the correction for critical density is omitted.  It has a steeper slope
than the observed distribution at low x values.  No physical
combination of free parameters resulted in a shallower slope than the
one shown by the diamonds.  Lowering $\alpha$ in the Schechter mass
equation to values that result in a divergent mass in galaxies lowered
the slope slightly but was considered non-physical.  The discussions in
\citet{ken89} and \citet{mar01} indicate that there is a critical
density that depends on dynamical factors through the Toomre Q factor
and possibly the amount of shear in the galactic rotation.  The form of
the reduction of star formation efficiency utilized in
eqn.~\ref{eqn:crit} is a simple way of reducing efficiency at low
surface density, i.e.  low x, regions.  The range of critical densities
discussed in \citet{mar01} and the observed fact that there is star
formation in regions that are below critical densities indicated a
softer reduction of the star formation efficiency than a simple
truncation.  The value of the exponent (-1.5) produces a good match to
the observed slope.  That such a reduction is required to match the
observations is further evidence for a reduction of star formation
efficiency below the rate predicted by the simple Schmidt law at low
surface densities.

The soft roll off of the Schmidt Law used in this work does not imply
that there is not a local critical density below which star formation
does not occur.  Even though the average density in the roughly 1
square kiloparsec area covered by a single pixel may be below the
critical density, there may be several areas where the local density is
higher than the critical density.  This is probably the best physical
interpretation of the soft roll off on the kpc scale.

\section{Evolution with Redshift}

The main parameters that control the shape of the distribution function
are $r_e$ and $m^*$.  The fit is relatively insensitive to $\alpha$ and
the value of $\phi^*$ mainly scales the distribution rather than change
its shape.  In hierarchical models of galaxy formation and evolution
both the value of $m^*$ and $r_e$ will decrease as the redshift
increases.  Lowering $m^*$ decreases the value of x where the
transition from power law to exponential occurs.  The effect of
reducing $r_e$ is just the opposite, it increases the value of x where
the transition occurs.  Both of these effects make physical sense.  The
exact evolution of the intensity distribution will then depend on the
evolution of these two parameters.  In an attempt to determine the
effect of evolution the computation was performed for an epoch where
the characteristic mass $m^*$ is 1/10 of the $3 \times 10^{10}$
M$_{\odot}$ found from the matching at z = 1.  The value of $r_e$ was
then reduced to  $(1/10)^{1/5}$ of its value at z=1.  The result is the
dashed line in fig.~\ref{fig:hx}.  The main effect is a transition to
the exponential function at a lower of x.  If we reduce $r_e$ by a
larger amount a distribution very close to the z = 1 distribution is
obtained.  The HST observations at redshifts greater than 1.5 do not
adequately define the shape of h(x) to provide an observational test of
the evolution of h(x).  Galaxy evolution modeler's predictions of the
free parameters and the resulting h(x) can be compared at high redshift
when NGST becomes operational.

\section{Conclusions}

We have shown that the star formation intensity distribution shape is a natural
consequence of a Schechter galaxy mass distribution, the Schmidt law
with a critical density and star formation occurring in exponential disks.
The primary parameters that control the shape of the distribution are the value
of $m^*$ in the Schechter function and the value of $r_e$ in the exponential
surface density distribution.  In a hierarchical galaxy formation scenario both
of these values would expect to be reduced at higher redshifts.  Reduction of
the values of these two parameters have opposite effects on the shape of the
function and may cancel each other out in part.

\acknowledgments

The author would like to acknowledge the very helpful comments of an
anonymous referee. This work is supported in part by NASA grant NAG
5-10843.  This work utilized observations with the NASA/ESA Hubble
Space Telescope, obtained at the Space Telescope Science Institute,
which is operated by the Association of Universities for Research in
Astronomy under NASA contract NAS5-26555.

\clearpage

\end{document}